\documentclass[11pt]{article}

\usepackage[margin=1in]{geometry}
\usepackage{setspace}
\usepackage{amsmath, amssymb}
\usepackage{hyperref}
\usepackage{natbib}
\usepackage{booktabs}
\usepackage{threeparttable}
\usepackage{graphicx}
\usepackage{float}

\hypersetup{
  colorlinks=true,
  linkcolor=blue,
  citecolor=blue,
  urlcolor=blue
}

\title{\vspace{-1.0cm}
Returns to U.S. and Foreign Experience among Immigrant Men:\\
Evidence from IPUMS Microdata
}
\author{Farhad Vasheghanifarahani\\
University of South Florida, Department of Economics}
\date{\today}

\begin{document}
\maketitle

\begin{abstract}
This paper studies wage returns to labor-market experience with a focus on immigrants and the portability of foreign-acquired human capital. Using a cleaned IPUMS microdata sample of male, full-time, private-sector workers, I describe cross-origin differences in normalized wages and experience profiles and estimate Mincer-style wage regressions with experience-group indicators and fixed effects. Descriptive evidence suggests that immigrants earn less than comparable nonmigrants within the same year on average, and that normalized wages rise with U.S. experience. Regression results indicate strong, increasing associations between wages and total experience in the pooled sample; however, experience coefficients are smaller when restricting to immigrants. A key finding is that the estimated returns to U.S. experience are large and monotonic across bins, whereas the returns to foreign experience are substantially smaller for most bins. A complementary country analysis for recent migrants from Canada, England, Mexico, and Guatemala shows steeper experience profiles for migrants from higher-income origin countries. Together, the results are consistent with imperfect transferability of foreign work experience and the central role of host-country human capital in immigrant wage growth.
\end{abstract}

\vspace{0.25cm}

\section{Introduction}

A standard starting point for measuring how human capital maps into earnings is the Mincerian earnings function, which relates (log) wages to schooling and labor-market experience \citep{Mincer1974}. While the Mincer framework is often introduced as a compact empirical approximation, it has also motivated a large theoretical and econometric literature clarifying what estimated coefficients capture, when they can be interpreted as returns to human capital, and how life-cycle dynamics and selection shape earnings profiles \citep{HeckmanLochnerTodd2006}. In practice, researchers frequently operationalize experience using ``potential experience'' (typically age minus years of schooling minus six), which is available in large cross-sections and provides a workable proxy for actual labor-market experience.

This paper studies \emph{returns to experience} using U.S. Census and ACS microdata and asks a closely related set of questions that are central to labor economics and international migration: do migrants receive the same returns to experience as non-migrants, and—among migrants—does experience accumulated in the United States command different payoffs than experience accumulated abroad? The distinction matters because experience is not merely time spent working; it is a bundle of skills and knowledge that can be country-specific. If skills are imperfectly transferable, immigrants may initially earn less than observationally similar natives even when they bring substantial prior work experience, and their wage growth may depend on how quickly they accumulate host-country human capital.

A large literature documents that immigrants' earnings tend to rise with time spent in the destination country, consistent with human capital accumulation after arrival \citep{Chiswick1978}. At the same time, the classic assimilation profile is not mechanically identified from a single cross-section: differences in earnings by years-since-migration can reflect true assimilation, but they can also reflect differences in cohort composition or immigrant ``quality'' across arrival waves \citep{Borjas1985,Borjas1995}. Influential syntheses emphasize that distinguishing assimilation from cohort effects requires careful comparisons across cohorts and over time, and that repeated cross-sections can sometimes exaggerate earnings growth relative to longitudinal evidence \citep{LaLondeTopel1992,Lubotsky2007}. These debates motivate specifications that flexibly control for time effects and compare groups in consistent ways across the sample period.

A complementary strand focuses on \emph{portability} of human capital: education and experience acquired abroad may be valued differently than human capital acquired in the destination \citep{Friedberg2000}. For example, immigrants who obtain host-country schooling often earn higher wages than otherwise similar immigrants educated only abroad, suggesting that destination-specific credentials and skills can be rewarded more strongly \citep{BratsbergRagan2002}. Other work emphasizes that immigrants may initially accept jobs below their formal skill level (occupational downgrading), and that mismatch can be particularly salient among some origin groups and among highly educated migrants \citep{Akresh2006,MattooNeaguOzden2008}. These findings underscore why it is useful to distinguish \emph{where} experience was accumulated and to allow returns to experience to vary across groups.

Language proficiency is another central mechanism linking migration to earnings. English skills can increase earnings directly by raising productivity and expand job opportunities, but language measures may also be endogenous to labor-market outcomes and correlated with unobserved ability or motivation \citep{ChiswickMiller1995,DustmannFabbri2003}. Empirical strategies exploiting age-at-arrival patterns have been used to sharpen identification of language effects on earnings, highlighting economically meaningful returns to language proficiency \citep{BleakleyChin2004}. In applied work using large cross-sections, incorporating flexible controls for English proficiency is therefore an important step for interpreting estimated experience profiles for migrants.

Guided by these literatures, I estimate Mincer-style wage regressions in which experience enters as a set of indicator groups (e.g., 0--5, 6--10, \dots, 36--40 years), allowing for flexible nonlinearity in the experience-earnings relationship. I then compare specifications that (i) pool migrants and non-migrants, (ii) add controls for English proficiency, (iii) incorporate state fixed effects, (iv) restrict to migrants only, (v) focus on ``new'' migrants, and (vi) decompose experience into potential U.S. experience versus potential foreign experience. This structure is designed to connect closely to the portability and assimilation mechanisms emphasized in prior work, while remaining transparent and replicable with Census/ACS microdata.

The remainder of the paper is organized as follows. Section~2 describes the data construction and sample restrictions. Section~3 outlines the empirical specifications. Section~4 presents the main regression results and discusses heterogeneity across origin countries and across definitions of experience. Section~5 discusses interpretation and limitations, and Section~6 concludes.


\section{Data and Sample Construction}\label{sec:data}

\subsection{Data source}
I use individual-level U.S. microdata from IPUMS (Integrated Public Use Microdata Series). The analysis pools repeated cross-sections and leverages harmonized demographic, labor market, and migration variables that are consistently defined across years. The empirical focus is on immigrant men, but U.S.-born men are retained as a comparison group for wage normalization.

\subsection{Sample restrictions}
I begin with working-age male individuals and impose a set of restrictions designed to approximate a relatively stable ``full-time private-sector'' workforce. Specifically, I keep men older than 17, restrict to private-sector workers (based on \texttt{CLASSWKR}), and define full-time status as working at least 35 usual hours per week (\texttt{UHRSWORK} $\geq 35$). Observations with missing values in the core variables used in the analysis are dropped so that the resulting dataset is internally consistent across the descriptive and regression exercises.

\subsection{Key variable construction}

\paragraph{Immigrant status and English proficiency.}
I define an indicator for immigrant status using reported year of immigration. An individual is classified as an immigrant if \texttt{YRIMMIG} is observed and indicates a valid year of entry.\footnote{In the raw extraction, some values correspond to ``not in universe'' or otherwise invalid codes; these are excluded from the immigrant definition.} For language, I use the IPUMS English-speaking measure (\texttt{SPEAKENG}) for immigrants. For non-immigrants, \texttt{SPEAKENG} is set to a distinct baseline value so that language controls apply naturally to immigrants while keeping the full sample intact for year fixed effects and normalization.

\paragraph{Years of schooling.}
I construct a proxy for years of schooling, \texttt{yrschool}, from the categorical education variable (\texttt{EDUC}). The mapping converts each education category to an equivalent number of completed years (e.g., high school completion mapped to 12, college completion mapped to 16, and so on). In a subset of lower-education categories, I use the more detailed education measure (\texttt{EDUCD}) to refine the schooling proxy when it is available.\footnote{This follows the coding approach used in the original problem-set workflow: for some low-education categories, \texttt{yrschool} is replaced by \texttt{EDUCD} when \texttt{EDUCD} is non-missing. Because \texttt{EDUCD} is more granular than \texttt{EDUC}, this adjustment can reduce coarseness at the bottom of the schooling distribution.}

\paragraph{Hourly wage and top-code adjustment.}
Let \texttt{INCWAGE} denote annual wage and salary income, \texttt{UHRSWORK} usual weekly hours worked, and \texttt{WKSWORK2} weeks worked in the prior year. I construct an hourly wage measure by dividing annual earnings by annual hours:
\begin{equation}
w_{it} \;=\; \frac{\texttt{INCWAGE}_{it}}{\texttt{UHRSWORK}_{it}\times \texttt{WKSWORK2}_{it}}.
\label{eq:wage}
\end{equation}
When \footnote{In the data cleaning, the 1980 top-coded value is inflated by a factor of 1.4 (i.e., the top code \texttt{INCWAGE}=\$75{,}000 in 1980 is recoded to \$75{,}000 $\times$ 1.4) before constructing hourly wages.}

\paragraph{Potential experience and decomposition into U.S.\ vs.\ foreign experience.}
I construct potential labor market experience using the standard Mincer-style proxy based on age and schooling:
\begin{equation}
\texttt{potexp}_{it} \;=\;
\begin{cases}
\texttt{AGE}_{it}-18, & \text{if } \texttt{yrschool}_{it} < 12,\\[4pt]
\texttt{AGE}_{it}-\texttt{yrschool}_{it}-6, & \text{otherwise.}
\end{cases}
\label{eq:potexp}
\end{equation}
For immigrants, I then decompose potential experience into experience accumulated in the United States and experience accumulated prior to arrival. U.S.\ potential experience is defined as time since immigration:
\begin{equation}
\texttt{potUSexp}_{it} \;=\; \texttt{YEAR}_{it} - \texttt{YRIMMIG}_{it},
\label{eq:potUSexp}
\end{equation}
and foreign potential experience is the residual:
\begin{equation}
\texttt{potFRexp}_{it} \;=\; \texttt{potexp}_{it}-\texttt{potUSexp}_{it}.
\label{eq:potFRexp}
\end{equation}
For non-immigrants, I set $\texttt{potUSexp}_{it}=\texttt{potexp}_{it}$ and $\texttt{potFRexp}_{it}=0$.

To keep the analysis within a common support and reduce leverage from extreme values, I restrict the sample to individuals with $\texttt{potexp}\in[0,40]$. Finally, I compute age at migration and impose an additional restriction intended to ensure schooling is plausibly completed prior to migration:
\begin{equation}
\texttt{age\_at\_migration}_{it} \;=\; \texttt{YRIMMIG}_{it} - (\texttt{YEAR}_{it}-\texttt{AGE}_{it}),
\end{equation}
and I retain immigrants with $\texttt{age\_at\_migration}>\texttt{yrschool}$.

\paragraph{Wage normalization.}
Because wage levels vary across years for reasons unrelated to assimilation (inflation, business cycles, and secular changes), I also construct a normalized wage outcome. For each year, I compute the average hourly wage among non-immigrants with positive wages and define:
\begin{equation}
\texttt{normwage}_{it} \;=\; \frac{w_{it}}{\overline{w}^{\,\text{non-imm}}_{t}}.
\label{eq:normwage}
\end{equation}
This normalization expresses individual wages relative to the contemporaneous non-immigrant benchmark and is useful for descriptive comparisons across cohorts and origin groups.

\subsection{Implementation and reproducibility}
All cleaning steps and empirical analysis are scripted in \texttt{R}, using a transparent pipeline that reads the raw IPUMS extract, constructs the analysis variables, and exports an analytic file used for tables and figures. This reproducible workflow mirrors best practices I have used in other applied projects that emphasize careful data processing and replicable modeling.\footnote{For an applied example of a fully scripted data-to-results pipeline (in a different context), see \citet{FiroozabadiAnsariVasheghanifarahani2024}.}


\section{Empirical Strategy}\label{sec:empirical}

\subsection{Baseline wage specification}
To quantify returns to experience, I estimate Mincer-style wage regressions where the dependent variable is the logarithm of hourly wage, $\log(w_{it})$. Because the relationship between wages and experience is often nonlinear, I discretize potential experience into five-year bins. Let $G_{it}\in\{0\text{--}5,\,6\text{--}10,\,\ldots,\,36\text{--}40\}$ denote the experience-group indicator constructed from potential experience. Using the $0$--$5$ group as the omitted category, the baseline specification is:
\begin{equation}
\log(w_{it}) = \beta_s \, \text{School}_{it}
+ \sum_{g\neq 0\text{--}5}\beta_g \,\mathbf{1}\{G_{it}=g\}
+ \delta_t
+ \varepsilon_{it},
\label{eq:baseline}
\end{equation}
where $\text{School}_{it}$ is years of schooling and $\delta_t$ are year fixed effects. Coefficients $\beta_g$ therefore measure the difference in log wages for group $g$ relative to the $0$--$5$ years-of-experience group, holding schooling and year effects constant.

\subsection{Accounting for English proficiency and local labor markets}
Language is an important component of immigrants' human capital. I therefore extend the baseline model by adding a set of English proficiency indicators. Let $L_{it}$ denote the English-speaking category (with a baseline category omitted). The augmented model is:
\begin{equation}
\log(w_{it}) = \beta_s \, \text{School}_{it}
+ \sum_{g\neq 0\text{--}5}\beta_g \,\mathbf{1}\{G_{it}=g\}
+ \sum_{\ell \neq \ell_0}\gamma_{\ell}\,\mathbf{1}\{L_{it}=\ell\}
+ \delta_t
+ \varepsilon_{it}.
\label{eq:english}
\end{equation}

Finally, to absorb time-invariant differences in wage levels across states (e.g., persistent cost-of-living, industrial composition, and institutional factors), I estimate a specification that includes state fixed effects:
\begin{equation}
\log(w_{it}) = \beta_s \, \text{School}_{it}
+ \sum_{g\neq 0\text{--}5}\beta_g \,\mathbf{1}\{G_{it}=g\}
+ \sum_{\ell \neq \ell_0}\gamma_{\ell}\,\mathbf{1}\{L_{it}=\ell\}
+ \delta_t + \mu_s
+ \varepsilon_{it},
\label{eq:statefe}
\end{equation}
where $\mu_s$ denotes state fixed effects.

\subsection{Migrant-only and ``new migrant'' subsamples}
To focus on immigrant assimilation patterns more directly, I re-estimate the model on an immigrant-only sample. In addition, I define a ``new migrant'' subsample based on time since arrival. Specifically, an observation is classified as a new migrant if the years since immigration is at most one year, with a wider window for 1980 and 1990 (up to five years since immigration) to ensure enough observations for stable estimation in those early census years. The migrant-only and new-migrant specifications retain year and state fixed effects and the same controls described above.

\subsection{Separating U.S.\ and foreign experience}
A central goal of the analysis is to compare the returns to experience acquired in the United States versus experience acquired abroad. To do so, I construct two experience-group variables: one based on U.S.\ potential experience and another based on foreign potential experience. Let $G^{US}_{it}$ denote the five-year bin for U.S.\ experience and $G^{FR}_{it}$ denote the five-year bin for foreign experience (each with $0$--$5$ as the omitted category). The decomposition model is:
\begin{equation}
\log(w_{it}) = \beta_s \, \text{School}_{it}
+ \sum_{g\neq 0\text{--}5}\beta^{US}_g \,\mathbf{1}\{G^{US}_{it}=g\}
+ \sum_{g\neq 0\text{--}5}\beta^{FR}_g \,\mathbf{1}\{G^{FR}_{it}=g\}
+ \sum_{\ell \neq \ell_0}\gamma_{\ell}\,\mathbf{1}\{L_{it}=\ell\}
+ \delta_t + \mu_s
+ \varepsilon_{it}.
\label{eq:usforeign}
\end{equation}
This specification allows U.S.\ experience and foreign experience to have different wage gradients, holding constant schooling, English proficiency, year effects, and state effects.

\subsection{Inference}
All models are estimated by OLS with fixed effects. Following the problem-set implementation, standard errors are clustered at the year level to allow for arbitrary correlation in the regression residuals within a survey year.

\clearpage
\section{Results}\label{sec:results}

\subsection{Descriptive Statistics}\label{subsec:desc}

Table~\ref{tab:desc} reports summary statistics for the immigrant sample used in the analysis. The mean normalized wage is 0.776, implying that (on average) immigrants in the analytic sample earn about 22 percent less than the average non-immigrant worker in the same year (by construction, the non-immigrant benchmark average equals 1). Immigrants have an average of 14.09 years of potential U.S. experience and 4.66 years of potential foreign experience. Average schooling is 16.40 years, and average English proficiency is 4.33 on the sample scale.

There is meaningful heterogeneity across origin countries. For instance, immigrants from the Philippines have a higher mean normalized wage (0.967) than immigrants from Mexico (0.634). China shows relatively high foreign experience (8.87) but comparatively lower normalized wages (0.679), consistent with the possibility that foreign experience is less transferable to the U.S. labor market than experience accumulated in the United States.

\begin{table}[H]
\centering
\caption{Descriptive Statistics (Immigrant Sample)}
\label{tab:desc}
\begin{threeparttable}
\resizebox{\textwidth}{!}{%
\begin{tabular}{lcccccc}
\toprule
Category & normwage & potUSexp & potFRexp & yrschool & speakeng & N \\
\midrule
Mean        & 0.7760 & 14.09 & 4.66 & 16.40 & 4.33 & 103,171 \\
S.D.        & 0.8115 & 9.57  & 8.24 & 4.89  & 1.64 & --- \\
\addlinespace
Vietnam     & 0.8664 & 17.19 & 6.24 & 16.47 & 5.05 & 3,387 \\
Mexico      & 0.6339 & 13.47 & 3.00 & 15.82 & 4.30 & 42,899 \\
El Salvador & 0.6856 & 14.50 & 3.14 & 15.46 & 4.43 & 4,237 \\
Philippines & 0.9673 & 17.23 & 5.95 & 17.51 & 4.56 & 3,562 \\
India       & 0.8738 & 13.68 & 7.95 & 17.72 & 4.57 & 1,460 \\
China       & 0.6789 & 14.19 & 8.87 & 17.31 & 4.27 & 2,203 \\
\bottomrule
\end{tabular}
}
\begin{tablenotes}[flushleft]
\footnotesize
\item \textit{Notes:} Sample includes male, full-time ($\geq 35$ usual hours), private-sector workers older than 17. Normalized wage equals hourly wage divided by the mean hourly wage of non-immigrants in the same year. \texttt{potUSexp} is potential U.S. experience (years since immigration); \texttt{potFRexp} is potential foreign experience. See Section~\ref{sec:data} for full variable construction.
\end{tablenotes}
\end{threeparttable}
\end{table}


\subsection{Experience Profiles for Recent Migrants by Origin}\label{subsec:fig1}

To illustrate heterogeneity in wage--experience profiles across origin countries near arrival, I estimate experience-group effects separately for recent migrants from Canada, England, Mexico, and Guatemala. ``Recent'' migrants are defined as those arriving within one year of the survey year, with a wider window (up to five years since arrival) in the 1980 and 1990 census years to ensure adequate sample sizes in early waves.\footnote{This definition follows the problem-set implementation used to construct the country-specific experience profiles.}

\begin{figure}[htbp]
\centering
\includegraphics[width=0.95\textwidth]{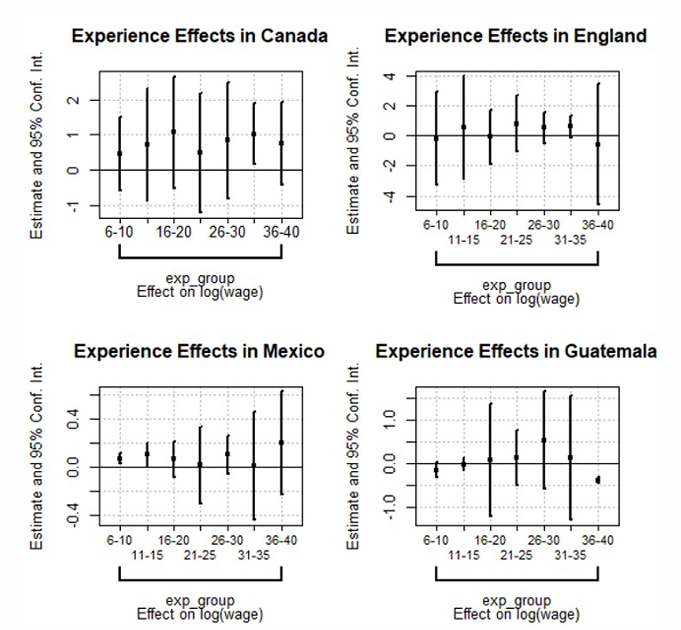}
\caption{Estimated experience effects for recent migrants from Canada, England, Mexico, and Guatemala. Points are experience-group coefficients from year fixed-effects regressions (relative to 0--5 years); bars show 95\% confidence intervals.}
\label{fig:countryexp}
\end{figure}

Figure~\ref{fig:countryexp} plots the estimated coefficients on experience-group indicators (relative to the 0--5 years group) and the associated 95\% confidence intervals. The experience gradient is steepest for Canada and England and flatter for Mexico and Guatemala. This pattern is consistent with the hypothesis that the portability of human capital differs across origin countries, including due to language and institutional proximity, which can affect how quickly migrants' skills translate into higher wages after arrival.

%

\subsection{Main Regression Evidence}\label{subsec:mainregs}

Table~\ref{tab:mainregs} reports the main Mincer-style regressions described in Section~\ref{sec:empirical}. Column (1) estimates the baseline model with schooling, experience-group indicators, and year fixed effects. The estimated coefficient on schooling is 0.0114, implying approximately a 1.1 percent higher hourly wage per additional year of schooling in this specification. Experience coefficients increase with experience, reaching about 0.43 log points for workers with 26--30 and 31--35 years of experience relative to the 0--5 years group.

Column (2) adds English proficiency controls. The schooling coefficient falls to 0.0089, consistent with English proficiency capturing an important component of human capital correlated with schooling in the pooled sample. Column (3) further adds state fixed effects, and the estimated experience profile remains stable, indicating that the baseline wage--experience gradient is not driven solely by persistent cross-state differences in wage levels.

Columns (4) and (5) restrict attention to immigrants. In the immigrant-only sample (column 4), the experience coefficients are smaller than in the pooled sample. For example, the coefficient for 26--30 years of experience is 0.294, compared with 0.423 in column (3). Focusing on new migrants (column 5), estimated experience effects remain positive but are less precisely estimated because the sample is much smaller.

Finally, column (6) decomposes experience into U.S.\ and foreign components. The estimated returns to U.S.\ experience are large and monotonic across bins, whereas the estimated coefficients on foreign experience bins are substantially smaller across most of the distribution. This contrast is consistent with imperfect transferability of foreign-acquired experience and a central role for host-country experience in immigrant wage growth.

\begin{table}[htbp]
\centering
\caption{Wage regressions with experience groups and U.S.\ vs.\ foreign experience}
\label{tab:mainregs}

\resizebox{\textwidth}{!}{%
\begin{tabular}{lcccccc}
\hline
 & (1) Baseline & (2) + English & (3) + State FE & (4) Migrants Only & (5) New Migrants & (6) US vs Foreign \\
\hline
\textbf{Schooling} & & & & & & \\
yrschool
& 0.0114*** & 0.0089*** & 0.0092*** & 0.0071** & 0.0083** & 0.0093*** \\
& (0.0003)  & (0.0003)  & (0.0003)  & (0.0009) & (0.0011) & (0.0003) \\
\hline
\textbf{Experience groups (omitted: 0--5)} & & & & & & \\
6--10  & 0.1975*** & 0.2034*** & 0.2064*** & 0.0979*** & 0.0741*** & --- \\
       & (0.0032)  & (0.0028)  & (0.0032)  & (0.0047)  & (0.0069)  &     \\
11--15 & 0.3050*** & 0.3097*** & 0.3113*** & 0.1994*** & 0.1027**  & --- \\
       & (0.0025)  & (0.0033)  & (0.0037)  & (0.0047)  & (0.0106)  &     \\
16--20 & 0.3701*** & 0.3708*** & 0.3689*** & 0.2526*** & 0.1638**  & --- \\
       & (0.0027)  & (0.0032)  & (0.0032)  & (0.0052)  & (0.0176)  &     \\
21--25 & 0.4155*** & 0.4109*** & 0.4082*** & 0.2860*** & 0.1900**  & --- \\
       & (0.0042)  & (0.0045)  & (0.0043)  & (0.0082)  & (0.0329)  &     \\
26--30 & 0.4315*** & 0.4244*** & 0.4231*** & 0.2943*** & 0.1970*** & --- \\
       & (0.0048)  & (0.0047)  & (0.0047)  & (0.0095)  & (0.0109)  &     \\
31--35 & 0.4320*** & 0.4257*** & 0.4258*** & 0.3057*** & 0.1333    & --- \\
       & (0.0058)  & (0.0051)  & (0.0051)  & (0.0106)  & (0.0486)  &     \\
36--40 & 0.4199*** & 0.4148*** & 0.4160*** & 0.3022*** & 0.1269    & --- \\
       & (0.0062)  & (0.0069)  & (0.0068)  & (0.0159)  & (0.1572)  &     \\
\hline
\textbf{English proficiency indicators} & & & & & & \\
SPEAKENG = 0 & --- & 0.5502*** & 0.5918*** & 0.5518*** & --- & 0.5061*** \\
            &     & (0.0038)  & (0.0074)  & (0.0162)  &     & (0.0044)  \\
SPEAKENG = 3 & --- & 0.5412*** & 0.5392*** & 0.5518*** & 0.4764** & 0.5194*** \\
            &     & (0.0141)  & (0.0177)  & (0.0162)  & (0.0679) & (0.0142)  \\
SPEAKENG = 4 & --- & 0.4099*** & 0.4375*** & 0.4482*** & 0.4031** & 0.3872**  \\
            &     & (0.0086)  & (0.0105)  & (0.0108)  & (0.0426) & (0.0105)  \\
SPEAKENG = 5 & --- & 0.3380*** & 0.3223*** & 0.3367*** & 0.2532*  & 0.2667**  \\
            &     & (0.0087)  & (0.0111)  & (0.0100)  & (0.0270) & (0.0060)  \\
SPEAKENG = 6 & --- & 0.1379*** & 0.1292*** & 0.1378*** & 0.0402   & 0.0994*** \\
            &     & (0.0090)  & (0.0105)  & (0.0101)  & (0.0336) & (0.0097)  \\
\hline
\textbf{U.S.\ vs.\ foreign experience (Column 6; omitted: 0--5)} & & & & & & \\
US exp.\ 6--10   & --- & --- & --- & --- & --- & 0.2032*** \\
                &     &     &     &     &     & (0.0089) \\
US exp.\ 11--15  & --- & --- & --- & --- & --- & 0.2022*** \\
                &     &     &     &     &     & (0.0099) \\
US exp.\ 16--20  & --- & --- & --- & --- & --- & 0.3057*** \\
                &     &     &     &     &     & (0.0047) \\
US exp.\ 21--25  & --- & --- & --- & --- & --- & 0.4049*** \\
                &     &     &     &     &     & (0.0045) \\
US exp.\ 26--30  & --- & --- & --- & --- & --- & 0.4214*** \\
                &     &     &     &     &     & (0.0044) \\
US exp.\ 31--35  & --- & --- & --- & --- & --- & 0.4241*** \\
                &     &     &     &     &     & (0.0069) \\
US exp.\ 36--40  & --- & --- & --- & --- & --- & 0.4145*** \\
                &     &     &     &     &     & (0.0063) \\
Foreign exp.\ 6--10   & --- & --- & --- & --- & --- & 0.0351*** \\
Foreign exp.\ 11--15  & --- & --- & --- & --- & --- & 0.0486*** \\
Foreign exp.\ 16--20  & --- & --- & --- & --- & --- & 0.0939*** \\
Foreign exp.\ 21--25  & --- & --- & --- & --- & --- & 0.0994**  \\
Foreign exp.\ 26--30  & --- & --- & --- & --- & --- & 0.0926*   \\
Foreign exp.\ 31--35  & --- & --- & --- & --- & --- & 0.0969**  \\
Foreign exp.\ 36--40  & --- & --- & --- & --- & --- & 0.1515    \\
\hline
Year FE  & Yes & Yes & Yes & Yes & Yes & Yes \\
State FE & No  & No  & Yes & Yes & Yes & Yes \\
Observations & 1,007,491 & 1,007,491 & 1,007,491 & 103,166 & 5,007 & 973,578 \\
$R^{2}$      & 0.0565 & 0.0812 & 0.1039 & 0.1355 & 0.0925 & 0.0983 \\
\hline
\end{tabular}%
}

\vspace{0.2cm}
{\footnotesize \textit{Notes:} Dependent variable is $\log(\text{hourly wage})$. Standard errors are clustered by YEAR. Significance: *** $p<0.01$, ** $p<0.05$, * $p<0.1$. Experience groups are five-year bins with 0--5 as the omitted category. Column (6) replaces total experience with U.S.\ experience groups and foreign experience groups.}
\end{table}

\subsection{U.S.\ versus Foreign Experience: Evidence on Portability}\label{subsec:usvsforeign}

The decomposition in column (6) of Table~\ref{tab:mainregs} provides direct evidence on the portability of foreign-acquired experience. Holding schooling, English proficiency, year effects, and state fixed effects constant, the coefficients on U.S.\ experience bins are large and monotonic. For example, relative to the reference group (0--5 years of U.S.\ experience), the coefficient rises to 0.4049 for 21--25 years and remains above 0.41 through 36--40 years. These magnitudes imply substantial wage differences associated with additional time spent accumulating U.S.-specific labor-market experience.

In contrast, the estimated coefficients on foreign experience bins are substantially smaller across most of the distribution. The coefficient for 6--10 years of foreign experience is 0.0351, and even for 21--25 years of foreign experience the coefficient is 0.0994. While several foreign-experience coefficients are statistically different from zero, their magnitudes are modest compared with the U.S.\ experience coefficients. The gap between the two sets of coefficients is consistent with the view that foreign work experience is only partially transferable to U.S.\ employers, whereas U.S.\ experience more directly reflects destination-relevant skills, credentials, and job-match quality.

Interpretation should nonetheless be cautious. Experience is measured using \emph{potential} experience rather than actual labor-market tenure, and foreign experience is constructed residually as potential experience minus years since immigration. If schooling and work histories differ systematically across origin groups, or if immigrants experience intermittent labor-force participation around migration, the foreign-experience measure may contain measurement error that attenuates estimated returns. Despite these limitations, the qualitative pattern is robust across specifications: wage growth is much more strongly associated with U.S.\ experience than with foreign experience, reinforcing the central role of host-country human capital accumulation in immigrant earnings growth.


\section{Discussion and Limitations}\label{sec:discussion}

\subsection{Discussion of main findings}
The empirical results point to a consistent pattern: wages increase with experience, but the wage--experience gradient depends importantly on whether experience is accumulated in the United States or abroad. In the pooled sample, the experience-group coefficients rise steadily relative to the 0--5 years group, indicating a strong association between experience and wages. When the sample is restricted to immigrants, the experience gradient becomes smaller, suggesting that the experience--wage relationship differs meaningfully between immigrants and non-immigrants even after controlling for schooling and fixed effects.

The most informative evidence comes from the decomposition that separates potential experience into U.S.\ and foreign components. Holding constant schooling, English proficiency, year effects, and state fixed effects, the estimated coefficients on U.S.\ experience bins are large and monotonic, while the coefficients on foreign experience bins are comparatively modest. Put simply, the results suggest that what immigrants learn and accumulate in the U.S.\ labor market is rewarded more strongly than what they bring from abroad. This pattern is consistent with the human-capital portability hypothesis emphasized by \citet{Friedberg2000} and with classic assimilation frameworks in which immigrants gradually improve job matches and acquire destination-specific skills after arrival \citep{Chiswick1978,Borjas1985,Borjas1995}.

The country-specific evidence for recent migrants offers complementary intuition. The experience profiles for Canada and England are steeper than those for Mexico and Guatemala, which is consistent with the idea that institutional similarity, language proximity, and credential recognition may shape the speed at which immigrants translate experience into higher wages. While this analysis is not designed to isolate mechanisms, the patterns fit the broader view that assimilation is uneven across origin groups and closely tied to how transferable (and how legible) pre-migration human capital is in the U.S.\ labor market.

\subsection{Key limitations and interpretation}
Although the results are economically intuitive and internally consistent across specifications, they should be interpreted as descriptive evidence rather than causal estimates of assimilation. Several limitations are especially important.

\paragraph{Potential experience versus actual labor-market experience.}
The analysis relies on \emph{potential} experience, constructed from age and schooling. This proxy is widely used in earnings-function applications, but it inevitably differs from actual experience when individuals delay labor-force entry, experience unemployment spells, or have intermittent labor supply. These gaps may be particularly relevant around migration, when employment histories can be disrupted. Measurement error in experience can attenuate estimated experience effects and can affect comparisons between U.S.\ and foreign experience components.

\paragraph{Foreign experience is constructed residually.}
Foreign experience is computed as potential experience minus years since immigration. This decomposition is appealing conceptually, but it embeds any measurement error in both components. For example, if immigrants obtain additional schooling after arrival, or if time since immigration is not a close proxy for time actually worked in the United States, the residual foreign-experience measure becomes noisier. This can mechanically produce smaller estimated coefficients on foreign experience. For this reason, the decomposition should be read as evidence that foreign experience is \emph{less strongly associated} with wages than U.S.\ experience, rather than as a precise estimate of a structural ``return'' to foreign tenure.

\paragraph{Repeated cross-sections and cohort composition.}
The dataset pools repeated cross-sections rather than following individuals over time. Therefore, differences in wages by experience group (or by years since migration) reflect comparisons across different people, not within-person wage growth. A central concern in the assimilation literature is that cross-sectional profiles can combine true assimilation with cohort effects and selection \citep{Borjas1985,Borjas1995,LaLondeTopel1992}. Year fixed effects help absorb aggregate shocks, but they do not fully resolve cohort heterogeneity. The results should thus be interpreted as robust cross-sectional associations in large microdata.

\paragraph{Sample selection and external validity.}
The analytic sample is restricted to male, full-time, private-sector workers. This choice improves comparability and reduces variation driven by labor supply, but it limits the generalizability of the results. Experience profiles may differ for women, for part-time workers, or for public-sector employment. In addition, hourly wages are constructed from annual wage income and weeks/hours measures, which can introduce further measurement error.

\paragraph{English proficiency controls.}
English proficiency is included as a set of categorical controls, and it meaningfully affects estimated schooling and experience coefficients. However, language is likely endogenous: immigrants who earn more may invest more in English, and unobserved ability can be correlated with both language and earnings \citep{ChiswickMiller1995,BleakleyChin2004,DustmannFabbri2003}. Accordingly, the English coefficients should be interpreted as descriptive adjustments rather than causal language returns.

\subsection{Overall implication}
With these caveats in mind, the analysis supports a clear implication: immigrant wage differences are strongly related to the accumulation of U.S.-specific experience, while foreign-acquired experience appears to be rewarded more weakly. This pattern is consistent with imperfect transferability of human capital and suggests that the economic process of assimilation is closely linked to the acquisition of skills, credentials, and job matches that are recognized and valued in the U.S.\ labor market.


\end{document}